\documentstyle[12pt]{article}
\begin{document}
\topmargin 0pt
\oddsidemargin=-0.4truecm
\evensidemargin=-0.4truecm
\baselineskip=24pt
\setcounter{page}{1}
\begin{titlepage}
\vspace*{0.4cm}
\begin{center}
{\LARGE\bf
Spontaneous Family Symmetry Breaking And Fermion Mixing}
\vspace{0.8cm}

{\large\bf  Zhijian Tao}

\vspace*{0.8cm}

{\em Theory Division, Institute of High Energy Physics, Academia Sinica\\
\vspace*{-0.1cm}
Beijing 100039, China\\}
\end{center}
\vspace{.2truecm}

\begin{abstract}
We propose a theory to describe fermion mixing. The theory respects the
maximal abelian family symmetries which are spontaneously broken down at
a large scale. We find that quark mixing can be well described in this
theory. Some concrete models are constructed. The crucial point for
model building is to introduce some new particles which are SU(2) gauge
singlets and carry -1/3 electric charge. The approximate family symmetries
in our low energy effective theory are naturally achieved.
Some interesting phenomena in the effective theory,
especially with two Higgs-doublets,
are discussed.
\end{abstract}
\vspace{2cm}
\centerline{}
\vspace{.3cm}
\end{titlepage}
\renewcommand{\thefootnote}{\arabic{footnote}}
\setcounter{footnote}{0}
\newpage
In the standard model (SM) since all the
Yukawa couplings are  taken as free
parameters, it lacks of an explanation on the pattern of the fermion masses and
mixing, i.e. all the fermion masses and Kobayashi-Maskawa (KM) matrix are
inputs of the model. This is usually thought as a shortcoming of the SM
because of too many parameters. Moreover, though gauge interaction
respects a flavor symmetry $G=U(3)_Q\times U(3)_U\times U(3)_D\times U(3)_L
\times U(3)_E$ acting on quarks and leptons, where Q and L are SU(2)-doublet
quarks and leptons, U, D, E are singlet quarks and leptons respectively,
the non-vanishing Yukawa couplings break this symmetry explicitly.
Therefore it seems that in SM there exists a special flavor basis, we
call it physical flavor basis. This basis is only corresponding to a
set of particular form of Yukawa couplings, which would generate correct
fermion masses and mixing. Then
this arises a question of how to define a physical flavor basis of fermion
in SM. We think there are generally two possible ways. One may postulate
a particular form of Yukawa coupling, an example is democracy mechanism
\cite{Fri},
only the basis related to this particular form of Yukawa coupling is
physical flavor basis, since in fact changing a basis through G transformation
generally changes the form of Yukawa coupling too.
Another way is to impose a certain family symmetry on fermions.
A physical flavor basis can be singled out depending on the assignment of
quantum number of the symmetry group.
In this case the physical basis is specified first, while the pattern of
fermion masses and mixing  are determined only after the symmetry breaking.

The family symmetry group  could be abelian or
non-abelian. For three families SU(2) and SU(3) are the usual choices for
non-abelian symmetry group \cite{Wil}.
 In this work we would like to look into the abelian case. We use F, S and T
to denote the first, second and third family. Then the maximal abelian
flavor symmetry group we consider is $U(1)_F\times U(1)_S\times U(1)_T$
acting on quarks and leptons \cite{Hall},
we will call them $G_q$ and $G_l$ respectively.
The Yukawa interaction respecting
this symmetry can be written as
\begin{equation}
L_Y=f_a{\bar Q_a}U_a\tilde{\Phi}+f_a'{\bar Q_a}D_a\Phi+g_a{\bar L_a}E_a\Phi
+h.c.
\end{equation}
Here a is family index, $\Phi$ is standard Higgs-doublet.
If the symmetry is not broken Yukawa couplings, therefore fermion masses,
are all diagonal. So there is no mixing among different generations at this
stage.
Of course the reason for this is that different generations are assigned to
different kind of conserved charges. To obtain non-trivial quark mixing, which
is necessary to explain the observed KM matrix, the family symmetry must
be spontaneously broken down at some scale $V$. The purpose of this work
is to construct a theory, we will call it a underlying theory, which can
realize the correct quark mixing after the flavor symmetry breaking.

We believe there are many possibilities to construct a underlying theory
which can give rise to the correct results on fermion masses and mixing.
But here we constrain ourself within the following conditions as our
criteria to build our underlying theory. (i) we demand $V$ is larger than
weak-scale $v$, namely SM could be viewed as a low energy effective theory of
the underlying theory. And besides weak-scale $v$, $V$ is
the only physical scale below grand unification (GUT) scale in the theory.
(ii) There exists a hierarchy on tree level Yukawa
couplings, $f_3>>f_2>>f_1$, similar relations are
true for other terms. This requirement
simply reflects the fact of fermion mass hierarchy. So the third family is
defined as the heaviest one. (iii) Off-diagonal elements of the Yukawa
interaction are induced only at loop level.
This is probably true for quarks and
charged leptons but not necessarily true for neutrinos in all the cases.
As a result of this ansatz, one can easily have an order of magnitude
explanation on the quark mixing as we will see below.

With these assumptions, even without a concrete model, one is able
to obtain some qualitative features of the fermion mixing. After spontaneous
family symmetry breaking, the Yukawa couplings are no longer diagonal. The
non-diagonal couplings are induced due to symmetry breaking effects. In
principle there are infinite number of operators induced in low energy
effective interaction, but the dominant operators are those with the
lowest dimension due to the fact that $V$ is larger than weak scale $v$.
Those lowest dimension operators, which must be $SU(2)\times U(1)$
gauge invariant, can be written as
\begin{equation}
L_Y^{eff}=L_Y+\xi_{ab}{\bar Q_a}U_b\tilde{\Phi}+\xi'_{ab}{\bar Q_a}D_b\Phi+
\eta_{ab}{\bar L_a}E_b\Phi+\eta'_{ab}L^T_aL_b\Phi^T\Phi+h.c.
\end{equation}
Here a, b are family indices. $L_a^TL_b$ combines to SU(2)-singlet or triplet.

The off-diagonal elements of the couplings $\xi_{ab}$, $\xi'_{ab}$ and
$\eta_{ab}$ are responsible for quark and charged lepton mixing. The
fourth term is mostly responsible for generation of neutrino mass and mixing.
{}From dimension analysis and the assumption (iii), we estimate that
$\xi_{ab},~ \xi'_{ab} (a\not= b )\sim\displaystyle{\frac{f_3}{16\pi^2}}
\lambda_{ab}$. $\xi$ and $\xi'$ are dimensionless, and
$1/16\pi^2$ is the loop factor. $f_3$ is the Yukawa coupling of the
heaviest quark. For t quark as heavy as 170 GeV, $f_3$ should be
order of one. Of course this is only a rough estimation. One order of
magnitude larger or smaller should be reasonable. All  uncertainties and other
effects can be absorbed in $\lambda_{ab}$. As for the leptons there is
a similar result, i.e. $\eta_{ab} (a\not=b)\sim
\displaystyle{\frac{g_3}{16\pi^2}}
\lambda_{ab}'$ but with $g_3\sim 10^{-2}$, for  tau mass is around
1.7 GeV. However, one should also notice the difference between leptons
and quarks at this point. For example in SM quark mixing is non-trivial, but
the lepton mixing is trivial due to absence of right-handed neutrino. In
a more complicated model, as we will
see below, lepton mixing can occur generally.
The fourth term is a dimension five operator and it breaks
lepton number explicitly by two units. If lepton number is not broken then
$\eta_{ab}'=0$, otherwise $\eta'_{ab}\sim\displaystyle{\frac{g_3}{16\pi^2V}}
\lambda_{ab}$. This should be a natural result since we expect a small
neutrino mass for large $V$.

We denote the upper and  down quarks,
charged lepton and neutrino mass matrixes as
$M^U$, $M^D$, $M^L$ and $M^{\nu}$ respectively. The diagonal
elements of these matrixes are inputs in our theory. In practice we choose
$M^U_{33}\sim 170$ GeV, ${M^U_{22}}\sim 1.5$ GeV$>>M^U_{11}$;
$M^D_{33}\sim 5$ GeV, ${M^D_{22}}\sim 0.2$ GeV$>>M^D_{11}$ and
$M^L_{33}\sim 1.7$ GeV, ${M^L_{22}}\sim 0.1$ GeV$>>M^L_{11}$;
The mass matrixes are diagonalized by a set of bi-unitarity transformations
on left, right-handed quarks and leptons. If U and V are the unitarity
matrixes acting on $U_L$ and $D_L$, the KM matrix is defined as
$K=U^+V$. With a not unreasonable choice
of parameters:  $\lambda_{32}/16\pi^2\sim 10^{-3}$ and  $\lambda_{32}
\sim 6\lambda_{31}\sim 6\lambda_{21}$, the matrix elements of U and V are
evaluated as
\begin{equation}\begin{array}{ll}
U_{32}\sim M^U_{32}/M^U_{33}\sim 6M^U_{31}/M^U_{33}\sim 6U_{31}\sim 10^{-3} &
U_{21}\sim M^U_{21}/M^U_{22}\sim 2\times 10^{-2}\\
{} & {}\\
V_{32}\sim M^D_{32}/M^D_{33}\sim 6M^D_{31}/M^D_{33}\sim 6V_{31}
\sim 3\times 10^{-2} &
V_{21}\sim M^D_{21}/M^D_{22}\sim 0.2
\end{array}\end{equation}
{}From this one can see that down quark mixing is larger than the mixing of
up quark, and mixing between the third and the first two generations is
smaller than mixing of the first two generations.
This results in  correct KM matrix elements at a factor of 2 or 3 level.
And in this theory one can easily understand why $K_{32}$
is much smaller than $K_{21}$.  In fact all the results are due to the
fact that the third generation is much heavier than other two,
 up quarks are heavier than the corresponding down quarks
 and the
non-diagonal elements of mass matrixes are only generated at one loop level.
One interesting observation is that the loop effects which induce
off-diagonal Yukawa interaction for quarks and charged leptons could
be independent
of the flavor symmetry breaking scale. Therefore one is free to push $V$ very
high to satisfy other requirements. In contradiction with this is the neutrino
mass, which is inversely proportional to the scale $V$, namely
$m_{\nu}\sim 10^{-4}\lambda'v^2/V$ GeV. To get neutrino mass smaller
than a few eV, V must be larger than $10^7- 10^8$ GeV, if $\lambda'$ is not
very small, i.e. $\lambda'\sim 0.1-1$.

Spontaneous breaking of the global family symmetries will result in some
Nambu-Goldston bosons. These bosons couple to quarks, leptons or both.
There exists strong constrains on these couplings from cosmological and
astrophysical considerations. Generally the constrains imply a lower limit
on the scale of symmetry breaking $V$. In our theory since these couplings
are induced only from loop diagrams. Hence the lower limit on $V$ is
about two order of magnitudes smaller than  in other cases. For quark-
Goldston coupling the strongest constrains are from SN1987a \cite{Turner},
which imply
a  lower bound  $V>10^{8}$ GeV. For lepton-Goldston coupling the
most stringent restriction is due to the cooling of red giants \cite{Morgan}.
It requires
that the scale $V$ must be larger than about $10^7$ GeV.

We now consider explicit models which yield this low-energy effective theory.

Model I: \hskip 1pc To extend SM we introduce a group of scalar particles
$H_{ab}$ which are SU(2)-singlets and carry -1/3 electric charge. Under
SU(3) color transformation, they behave like $\bf{3}$  or $\bf{\bar 6}$ plets.
Scalars with these properties will couple to quarks as following:
\begin{equation}
L_Y'=g_{ab}Q_a^TQ_bH_{ab}+g_{ab}'U^T_aD_bH_{ab}+h.c.
\end{equation}
In order to break the family symmetry spontaneously, we introduce another set
of color and electric neutral particles $S_a$ which transform as singlet of
SU(2) gauge group. The most general Lagrangian of the model can be written
as
\begin{equation}
L=L_G+L_Y+L_Y'+V_P(S,H,\Phi)
\end{equation}
where $L_G$ is gauge interaction. V is the most general potential of the
scalar particles. Especially a term like $\lambda_{ac}H_{ab}^+H_{bc}S^*_aS_c$
is included in $V_P$.  In this model because the particles H and S do not
couple to leptons. We expect the lepton sector is same as that in SM. The
family symmetry on lepton sector
$G_l=U(1)_e\times U(1)_{\mu} \times U(1)_{\tau}$ is not
broken. As a result, $\eta_{ab}=\eta'_{ab}=0$ (see equ. (2)).
At scale $V$, when S fields get  nonzero vacuum expectation values, the
family symmetry $G_q$ (referring to quark fields only) is broken down to
$U(1)_B$, where B is baryon number. The quark mixing among
different generations are generated at one loop level as shown in Fig. 1.
It is straightforward to obtain
\begin{equation}\begin{array}{l}
\xi_{ab}\simeq\displaystyle{\frac{1}{16\pi^2}g^*_{ac}\lambda_{ab}(f_c'g_{bc}'
+f_bg_{bc})\frac{<S^*_a><S_b>}{m_H^2}}\\
\xi_{ab}'\simeq\displaystyle{\frac{1}{16\pi^2}g^*_{ac}\lambda_{ab}(f_cg_{bc}'
+f_b'g_{bc})\frac{<S^*_a><S_b>}{m_H^2}}\\
\end{array}\end{equation}
Here $m_H$ is the mass of H particle and it is understood one should
sum over index c. With $<S>\sim m_H\sim V$ and all the elements of
g and g$'$ at same order, we get an approximate form for $\xi$ and $\xi'$ as
\begin{equation}
\xi_{ab}\sim\xi_{ab}'\simeq\displaystyle{\frac{f_3}{16\pi^2}\lambda_{ab}}
\end{equation}
which is the form we discussed in  previous section. And obviously they are
independent of the scale $V$.

Model II: \hskip 1pc We add another Higgs-doublet in model I. This is a two-
Higgs-doublets model with abelian family symmetry $G_q$ spontaneously
broken down to $U(1)_B$. $G_l$ is still unbroken in this model. So we will
only consider quark sector. The low energy effective Yukawa interaction
after family symmetry breaking has the form
\begin{equation}
L_Y={\bar Q}(F_1\tilde{\Phi}_1+F_2\tilde{\Phi}_2)U+{\bar Q}(F_1'\Phi_1
+F_2'\Phi_2)D+h.c.
\end{equation}
Still we impose the hierarchy condition ( i.e. condition
(ii) ) on the diagonal elements of both $F_1(F_1')$ and $F_2(F_2')$
in order to obtain the pattern of quark masses naturally. The description
of quark mixing is similar as in model one. Since typically the off-diagonal
elements are about $0.2\sim 0.01$ of the related diagonal elements. We actually
obtain a two-Higgs-doublets model at low energy with an approximate family
symmetry $G_q$. One interesting consequence of the approximate symmetry is that
one is able to avoid too large flavor changing neutral current (FCNC)
naturally without resorting to Glashow and Weinberg's discrete symmetry
\cite{Glashow}.
Spontaneous CP violation is also possible in this model \cite{Lee}.
CP violation
phenomena of this model was discussed recently by Wu and Wolfenstein
\cite{Wu}, where
the approximate flavor symmetry is assumed from beginning. The main results
of their research are the $\epsilon'$ value originated from scalar interaction
is larger than in generic superweak model and in certain cases $\epsilon'/
\epsilon$ could be as large as $10^{-3}$. The electric dipole moments of
electron \cite{Barr1} and neutron \cite{Weinberg}
are close to current experimental bounds in this
kind of models. Here we would like to point out a very interesting
prediction of $D^0-{\bar D^0}$ mixing in our model. Since the mixing
matrices for both down and up quarks are determined approximately,
we can estimate the $D^0-{\bar D^0}$ mixing due to tree level FCNC. Our
estimation gives $\Delta m_D\sim 10^{-14}-10^{-13}(100GeV/m_H)^2$ GeV, which
is very close to present experimental bound \cite{Com},
while the tree level FCNC
contribution to $K^0-{\bar K^0}$ mixing is close but below the experimental
value. This is probably the best place to discover this new physics in
future experiments.

Model III: \hskip 1pc one group of SU(2)-singlet particles $h_{ab}$ are
introduced based on model II. $h_{ab}$ carry one units of electric charge.
These particles couple to lepton sector and the Higgs-doublets as
following
\begin{equation}
L_Y^{lepton}=C_{ab}L^T_aL_bh_{ab}+C'_{ab}\Phi_1^T\Phi_2h_{ab}S_{ab}
\end{equation}
Just for the lepton sector this model is well-known as Zee-type model
\cite{Zee}.
The lepton sector family symmetries $G_l$ are spontaneously broken at
scale $V$. $G_q$ and $G_l$ breaking are triggered by same S fields, therefore
they should be spontaneously broken at same scale. And the lepton
number violation in the model  implies neutrino will get non-zero mass.
This means nonzero $\xi$ and $\xi'$ values, $\xi\sim\xi'\sim\displaystyle{
\frac{g_3}{16\pi^2}\lambda}$.
In addition, there exists non-trivial mixing of charged leptons. We checked
all possible flavor changing processes in lepton sector
and found that those flavor
changing scalar interaction will not induce direct
observable effects in the current
and future experiments. Most phenomena of Zee-type model has been discussed
extensively in literatures \cite{Smirnov}.
Here we just point out that in our model
leptons and quarks  are actually treated in a similar way, i.e. we
introduce H particles to couple quark sector and at the same time h particles
to lepton sector.

In summary, we have examined the theory with $U(1)^3$ family symmetry which is
spontaneously broken down at a large
energy scale. We found that KM matrix can be well reproduced.
And at the same time an approximate abelian family symmetry at low energy
is automatically achieved. We constructed some explicit models by
introduction of the particles $H_{ab}$ in quark sector.
Analogy with this is the Zee-scalar particles $h_{ab}$  which was introduced
in lepton sector to generate neutrino mass. The quark mixing
could be independent of the family symmetry breaking scale and the mass of
H particle. Constrains from cosmology and astrophysics, and small neutrino
masses imply the flavor symmetry breaking scale should be larger than
$10^7\sim 10^8$ GeV. In models with two-Higgs doublets ( model II and III )
$D^0-{\bar D}^0$ mixing is expected to be large and very close to current
experimental bound \cite{Hall}.

Finally some further developments which we believe could be interesting are
in order. (1) In the framework of this theory, we don't have any idea on
the mass hierarchy of fermions. A larger unification symmetry of flavor is
probably needed. For instance, it is already known that in some models
with SU(3) flavor symmetry, one generation fermion could be "naturally"
much heavier than another two \cite{Barr2}. One interesting possibility
is to consider flavor symmetry breaking in two steps. At the first
step SU(3) symmetry is broken down to $U(1)^3$. This possibility is now under
investigation. (2) It is very desired to incorporate the models in
GUT theory. h particles can occur through 10 plets of SU(5). H
particles can be the number of, for example, 45 plets of Higgs fields in SU(5).
Though as we said we
can deal with quarks  and leptons in the similar way as in
model III,
it is not in any sense an unification theory.
A grand unification theory with certain flavor symmetry probably
will offer us a deeper understanding of the pattern of fermion mass and
mixing.


\begin{thebibliography}{1.}

\bibitem{Fri} H. Fritzsch and J. Plankl, Phys. Lett. {\bf B237}, 451 (1990);
Y. Nambu, in Proc. XI. Warsaw Symp. On High Energy Physics (Kazimierz,
Poland, 1988); P. Kuas and S. Meshkov, Mod. Phys. Lett. {\bf A3}, 1251 (1988).
\bibitem{Wil} D. Chang, P. Pal and G. Senjanovi\'c, Phys. Lett. {\bf B153},
407 (1985); G. Gelmini, S. Nussinov and T. Yanagida, Nucl. Phys. {\bf B219},
31 (1983); F. Wilczek, Phys. Rev. Lett. {\bf 49}, 1549 (1982);
F. Wilczek and A. Zee, Phys. Rev. Lett. {\bf 42}, 421 (1979).
\bibitem{Hall} This kind of approximate family symmetry has been discussed
recently. See , L. Hall and S. Weinberg, Phys. Rev. {\bf D48}, 979 (1993);
A. Antaramian, L. Hall and A. Rasin, Phys. Rev. Lett. {\bf 69}, 1871 (1992).
\bibitem{Turner} See for example, M. Turner, Phys. Rev. Lett. {\bf 60}, 1797
(1988).
\bibitem{Morgan} M. Morgan and G. Miller, Phys. Lett. {\bf 179B}, 379 (1986).
\bibitem{Glashow} S. Glashow and S. Weinberg, Phys. Rev. {\bf D15}, 1958
(1977).
\bibitem{Lee} T. D. Lee, Phys. Rev. {\bf D8}, 1226 (1973); Phys. Rep. {\bf 9C},
143 (1974).
\bibitem{Wu} Y.-L. Wu and L. Wolfenstein, Phys. Rev. Lett. {\bf 73}, 1762
(1994); ibid 2819 (1994); See also L. Hall and S. Weinberg in Ref. 3.
\bibitem{Barr1} S. M. Barr and A. Zee, Phys. Rev. Lett. {\bf 65}, 21 (1990).
\bibitem{Weinberg} S. Weinberg, Phys. Rev. {\bf D42}, 860 (1990); D. W. Chang,
W. Y. Keung and T. C. Yuan, Phys. Lett. {\bf B251}, 608 (1990); J. Gunion
and D. Wyler, Phys. Lett. {\bf B248}, 170 (1990)
\bibitem{Com} Hall and Weinberg obtained a similar result on $\Delta m_D$,
see the discussion on this issue in Ref. 3.
\bibitem{Zee} A. Zee, Phys. Lett. {\bf B93}, 387 (1980); {\bf B161},
141 (1985).
\bibitem{Smirnov} See for example, A. Yu. Smirnov and Z. Tao, Nucl .Phys. {\bf
B246}, 415 (1994);
R. Barbieri and L. Hall, Nucl. Phys. {\bf B364}, 27 (1991); S. M. Barr,
E. M. Freire and A. Zee, Phys. Rev. Lett. {\bf 65}, 2626 (1990); S.
T. Petcov, Phys. Lett. {\bf B115}, 401 (1982); L. Wolfenstein, Nucl. Phys. {\bf
B175}, 93 (1980).
\bibitem{Barr2} S. M. Barr, D. B. Reiss and A. Zee, Phys. Lett. {\bf B116},
227 (1982).
\end{thebibliography}
\end{document}